\documentclass[showpacs,preprintnumbers,amsmath,amssymb]{revtex4}

\usepackage{epsfig}

\newcommand\nn{\nonumber}
\newcommand\ba{\begin{eqnarray}}
\newcommand\ea{\end{eqnarray}}
\renewcommand{\Re}{\mbox{Re}}
\newcommand{\GeV}{~\mbox{GeV}}
\newcommand{\MeV}{~\mbox{MeV}}

\begin{document}

\title{Coulomb Corrections to the Parameters of the Moli\`{e}re
Multiple Scattering Theory }

\author{Eduard Kuraev}
\email[E-mail:]{kuraev@theor.jinr.ru}
\author{Olga Voskresenskaya}
\email[E-mail:]{voskr@jinr.ru}
\author{Alexander Tarasov}
\affiliation{Joint Institute for Nuclear Research, Joliot--Curie 6,
141980 Dubna, Moscow Region, Russian Federation}

\date{December 27}

\begin{abstract}
High-energy Coulomb corrections to the parameters of the Moli\`{e}re
multiple scattering theory are obtained. Numerical calculations are
presented in the range of the nuclear charge number of the target
atom $6\eqslantless Z\eqslantless 92$. It is shown that these
corrections have a large value for sufficiently heavy elements of
the target material and should be taken into account in describing
high-energy experiments with nuclear targets.
\end{abstract}


\maketitle

{\it Introduction.} The Coulomb correction (CC) is the
difference between the exact Born parameter $\xi$ result and
the Born approximation. At intermediate energies, formulas for the
Coulomb corrections are not available in analytical form
\cite{Koch}. The analytic formulas for the high-energy CC are known
as the Bethe-Bloch formulas for the ionization losses \cite{Bloch}
and those for the Bethe-Heitler cross section of bremsstrahlung
\cite{Heitler}.

A similar expression was found for the total cross section of the
Coulomb interaction of hadronic  atoms with ordinary target atoms
\cite{TVG}. Also, Coulomb corrections were  obtained to the cross
sections of the pair production in nuclear collisions
\cite{Heitler,Ivanov,oth}, a two-dimensional potential \cite{K}, and
the spectrum of bremsstrahlung \cite{Heitler,overbo}. The
specificity of the expressions presented in this work is that they
determine the Coulomb correction to some important parameters of the
Moli\`{e}re multiple scattering theory, i.e. the screening angular
parameter $\chi_a^{\scriptscriptstyle B}$ and also the parameters
$b$, $B$, and $\overline{\theta^2}$ of the Moli\`{e}re expansion
method \cite{M47}.

Moli\`{e}re's theory is of interest for numerous applications
related to particle transport in matter; is widely used  in most of
the transport codes; and also presents the most used tool for taking
into account the multiple scattering effects in experimental data
processing (the DIRAC experiment \cite{Dirac05} like many others
[11--13]).

As the Moli\`{e}re theory is currently used roughly for
$1\,\MeV$--$200\,\GeV$ proton beams \cite{others1,Strig} and
extremely high energy cosmic rays and can be applied to investigate
the IceCubes neutrino-induced showers \cite{OurIce} with energies
above 1 PeV \cite{Klein2}, the role of the high-energy CC to the
parameters of this theory becomes significant. Of especial
importance is the Coulomb correction to the screening angular
parameter,  as just this single parameter enters into other
important quantities of the Moli\`{e}re theory and describes the
scattering.

In his original paper, Moli\`{e}re obtained an approximate
semianalytical expression  for this parameter, valid to
second order in $\xi$, where only first term  is determined quite
accurately, while the coefficient in the second term is found
numerically and approximately.

In this work, we obtained for $\chi^{}_a$ and some other parameters
of the Moli\`{e}re theory exact analytical results valid to all
orders in $\xi$. We also evaluated numerically Coulomb corrections
to the Born approximation of these parameters accounting all orders
in $\xi$ over the range $6\eqslantless Z\eqslantless 92$.
Additionally, we estimated the accuracy of the Moli\`{e}re theory in
determining the screening angle. This Letter is organized as
follows: We start from the consideration of the standard approach to
the multiple scattering theory proposed by Moli\`{e}re. Then we
obtain the analytical and numerical results for the Coulomb
corrections to the parameters of the Moli\`{e}re theory. Finally, we
briefly summarize our findings.

{\it Moli\`ere multiple scattering theory.}  Multiple scattering
of charged high-energy particles in the Coulomb field of nuclei,
perpendicular to the incident particle direction, is a diffusion
process in the angular plane of $(\theta,\phi)=\vec{\chi}$. We
assume $\chi=\sin\theta\approx \theta \ll 1$; also we define
$\sigma(\chi)\chi \,d\chi \,d\phi$ as the differential cross section for
the single elastic scattering into the angular interval
$\chi,\chi+d\chi$. Define now $W(\theta,t)\theta d\theta$ as the
number of projectiles scattered in the angular interval $d\theta$
after traveling through an absorber of a thickness $t$ and its
normalization condition $\int W(\theta,t)\theta d \theta = 1$. For a
homogeneous absorber and fast charged particles, within the
small-angle approximation, the standard transport
equation can be used \cite{Scott,B}:
\ba \frac{\partial W(\theta,t)}{\partial
t}&=&-n\,W(\theta,t)\int\sigma(\chi)\chi d\chi\nn\\
     &&+ \;n\int
W(\vec\theta-\vec\chi,t)\sigma(\chi)d^2\chi \ .  \ea
Here $n$ is the density of the scattering centers per unit volume;
$d^2\vec{\chi}=\chi d\chi d\phi/(2\pi)$, and $\phi$ denotes the
azimuthal angle of the vector $\vec{\chi}$. The first term in the
right-hand side describes the decrease in the number of
projectiles from the cone $\theta$; and the second one, the
increase in the cone from the outside of the cone.

Following Moli\`{e}re \cite{M47,B}, we introduce the Bessel
transformation of distribution
\ba\label{eq3}
g(\eta,t)\;&=&\int\nolimits_0^\infty \theta J_0(\eta\theta)W(\theta,t)d\theta\ ,  \\
\label{W}W(\theta,t)&=&\int\nolimits_0^\infty \eta
J_0(\eta\theta)g(\eta,t)d\eta\ . \ea

For $g(\eta,t)$, using the folding theorem we obtain

\ba\label{eq5} \frac{\partial g(\eta,t)}{\partial t}=-n\,
g(\eta,t)\int\nolimits_0^\infty\sigma(\chi) \chi
d\chi[1-J_0(\eta\chi)]\  .\ea

Its solution is
\ba
\label{g(eta)}g(\eta,t)\;&=&\exp\left\{N(\eta,t)-N_0(0,t)\right\}\
, \\
N(\eta,t)&=&n\, t\int\sigma(\chi)\chi d\chi J_0(\eta\chi)\ . \ea

Inserting this expression in the Bessel transform of the
distribution function (\ref{W}), we get
\ba W(\theta,t)&=&\int\nolimits_0^\infty\eta d\eta
J_0(\eta\theta)\nn\\
\label{com} &&\times\exp\biggl\{-n\, t \,\int\nolimits_0^\infty
\sigma(\chi)\chi d\chi\left[1-J_0(\eta\chi)\right]\biggl\}. \ea

Let us write
\ba n\, t\,\sigma(\chi)\chi d\chi &=&2\chi_c^2\chi d\chi
q(\chi)/\chi^4\ ,\nn\\
\chi_c^2&=& 4\pi n\, tz^2Z(Z+1) e^4/(p v)^2\ . \ea
The quantity $q(\chi)$ is the ratio of the actual differential
scattering cross section to the Rutherford one; it describes the
deviation of the real potential from the Coulomb one. The Rutherford
scattering cross section is determined by
\ba \frac{d\sigma^{}_{\scriptscriptstyle R}(\chi)}{d
O}=\left(\frac{2zZ e^2}{mv^2}\right)^2\frac{1}{\chi^4},\ea
where $dO=\sin\theta d\theta d\phi$ represents the angular phase
volume, $e$ is the elementary charge, $m$ and $v$ are the mass of
the charged scattered particle and its velocity at large distances
from the scattering center which is assumed to be at rest, $r$ is
the distance between them,  $U(r)$ is the unscreened Coulomb
potential $U(r) = zZ e^2/r$, $z$ denotes the charge number of the
scattered particle, $e^2/\hbar c = 4\pi\alpha$, and $\alpha=1/137$
is the fine structure constant.

For the screened potential, the differential scattering cross
section reads
\ba \frac{d\sigma^{}(\chi)}{d O}=\left(\frac{2zZ
e^2}{pv}\right)^2\left(\frac{1}{\chi^2+\chi_0^2}\right)^2 \ ,\ea\\
where $\chi^{}_0=\hbar/pa$, $a=0.885\,a^{}_0Z^{-1/3}$, $p=mv$,
and the total cross for the single elastic scattering becomes
 \ba\label{sigma} \sigma = 2\pi \int\nolimits_0^\infty \sigma(\chi) \chi
d\chi =4\pi a^2\left(\frac{zZe}{\hbar v}\right)^2\ . \ea
Here $a^{}_0$ is the Bohr radius, and $a$ is the Fermi radius of the
target atom. If  the target thickness satisfies the condition $t \ll
l$, where $l = 1/(n \sigma)$, the distribution function can be
written as
\ba \label{W_e} W(\theta,t) = n t \sigma(\theta)\ ,\ea
where $\sigma(\theta)$ is the particle interaction cross section
with a separate scattering center. In this case it represents the
single-scattering probability. In the case when $t \gg l$, the
accounting of the multiple scattering is necessary, and the
distribution function  should be determined by (\ref{com}). For the
reasonable thickness, the width of the multiple scattering
distribution is very large compared with $\chi^{}_0$.

The quantity $q(\chi)$ is equal to unity for large values of
$\chi\geq \chi^{}_0$ and tends to zero at $\chi=0$. It contains
deviation from the Rutherford formulae due to the effects of
screening of atomic electrons and the Coulomb corrections arising
from multi-photon exchanges between the scattered particle and the
atomic nuclei. The main $\chi$ values belong to the region $\chi\sim
\chi^{}_0$.

The physical meaning of $\chi^{}_c$ can be understood from the
requirement that the probability of scattering on the angles
exceeding $\chi^{}_c$ is unity:
\ba n\, t\int\nolimits_{\chi_c}^\infty d\sigma(\chi)=\frac{4\pi n
t\, z^2Z(Z+1)e^4}{(pv)^2}
\int\nolimits_{\chi_c}^\infty\frac{d\chi}{\chi^3}=1\ . \ea
This formula is based on the Rutherford cross section and is the
definition of the angle $\chi^{}_c$. We replace $Z^2\to Z(Z+1)$
keeping in mind the scattering on atomic electrons. Below, we assume
that $z=1$.

Typically, $\chi^{}_c/\chi^{}_0\sim 100$. In terms of $\chi^{}_c$,
the solution of (\ref{eq5}) can be represented as follows
\cite{M47,B}:
\ba -\ln
g(\eta,t)&=&N_0(0,t)-N(\eta,t)\nn\\
&=&2\,\chi_c^2\int\nolimits_0^\infty\frac{d\chi}{\chi^3}\,q(\chi)\left[1-J_0(\chi
\eta)\right] . \ea
Introducing some quantity $k$ from the region $(\chi^{}_0,
\chi^{}_c)$, $\chi^{}_0\ll k\ll\chi^{}_c$, and considering the
contribution of the range $\chi<k$, Moli\`{e}re introduced the
notation of the screening angle $\chi^{}_a$
\ba -\ln\chi^{}_a=\lim_{k \to
\infty}\left[\int\nolimits_0^k\frac{d\chi}{\chi}\,q(\chi)+
\frac{1}{2}-\ln k\right]\ . \ea

One of the most important results of the Moli\`{e}re theory is that
the scattering is described by a single parameter, the  screening
angle $\chi^{}_a$ ($\chi_a^{\,\prime}$)
\begin{equation}\chi_a^{\,\prime}=\sqrt{1.167}\,\chi_a=
\left[\exp\left(C_{\scriptscriptstyle
\mathrm{E}}-0.5\right)\right]\chi_a\approx1.080\,\chi_a\ ,
\end{equation}
where $C_{\scriptscriptstyle \mathrm{E}}=0.577\ldots$ ~is the Euler
constant.

More precisely, the angular distribution depends only on the
logarithmic ratio $b$
\begin{equation}\label{b} b=\ln \left(\chi_c/\chi_a^{\,\prime} \right)^2\equiv\ln
\left(\chi_c/\chi_a \right)^2+1-2C_{\scriptscriptstyle E}
\end{equation}
of the characteristic angle $\chi_c$ describing the foil thickness
\begin{equation}\label{char} \chi_c^2=4\pi nt\left(Z\alpha/\beta p
\right)^2,\quad p=mv,\quad \beta=v/c\ , \end{equation}
to the screening angle $\chi_a^{\,\prime}$, which characterizes the
scattering atom.

In order to obtain a result valid for large angles,  Moli\`{e}re
defined a new parameter $B$ by the transcendental equation
\begin{equation}\label{B} B-\ln B=b\ . \end{equation}
The angular distribution function can be written then as
\begin{eqnarray} W(\theta,B) &=&\frac{1}{
\overline{\theta^2}}\int\nolimits_0^{\infty}y dy \,J_0 (\theta
y)e^{-y^2/4}\nn\\
&&\times\exp\left[\frac{y^2}{4B}\ln\left(\frac{y^2}{4}\right)\right],\quad
y=\chi_c\eta\ ,
\end{eqnarray}
where $\overline{\theta^2}$ is the mean  square scattering angle.

The Moli\`{e}re  expansion method is to consider the term
$y^2\ln(y^2/4)/4B$ as a small parameter.  This allows expansion of the
angular distribution function in a power series in $1/B$:
\begin{equation}\label{power}
W(\theta,t)=\sum\nolimits_{n=0}^{\infty}\frac{1}{n!}\frac{1}{B^n}W_n(\theta,t)\
,
\end{equation}
in which
\begin{equation*} W_n(\theta,t) =
\frac{1}{\overline{\theta^{\,2}}}\int\nolimits_0^{\infty}\!\!\!\!y
dy \,J_0 \!\left(\!\frac{\theta}{\sqrt{\overline{\theta^2}}}\,
y\!\right)
\!e^{-y^2/4}\!\left[\frac{y^2}{4}\ln\!\left(\frac{y^2}{4}\right)\right]^n
\!\!,
\end{equation*}
\begin{equation}\label{vartheta2}\overline{\theta^{\,2}}=\chi_c^2B=4\pi
nt\left(Z\alpha/\beta p \right)^2B(t)\ .
\end{equation}
This method is valid for $B\geq 4.5$ and
$\overline{\theta^{\,2}}<1$.

In order to obtain a result valid for large angles $\chi$ and also
for large $\xi=Z\alpha/\beta$, Moli\`{e}re used the WKB method and
a rather rough approximation in describing the screening angle:
\ba\label{1} \chi^{\scriptscriptstyle
M}_a&=&\chi_a^{\scriptscriptstyle B}\sqrt{1+3.34\,\xi^2}\ . \ea
This formula is determined only up to second order in $\xi$; its
coefficient in the second term  is found approximately using an
interpolation scheme.

Below we will use the eikonal approximation to obtain an exact
analytical expression for the Coulomb correction to the Born
screening angle $\chi_a^{\scriptscriptstyle
 B}=\sqrt{1.13}\chi^{}_0$. The accuracy of the eikonal approximation used
 below is the accuracy of the small-angle approximation \cite{BSh}, i.e.
$1+O\left(\chi_0/\chi_c\right)=1+O(10^{-2}),$ which is better than one
percent.

{\it Coulomb correction to the screening angular parameter.} Recall now
the relations for the scattering amplitude in the eikonal
approximation (see, e.g., \cite{BSh,BK}):
\ba f(\vec{q})&=&\frac{1}{2\pi i}\int d^2b\, \exp\left(\!-i
\vec{q}\,\vec{b}/\hbar\right)S(b)\ ,\nn\\
S(b) &=&\exp\left(\!-i
%\frac{
\phi(b)/\hbar\right)-1,\quad r=\sqrt{b^2+z^2}\\
\phi(b)&=&\frac{Ze^2}{\beta}\int\nolimits_{-\infty}^\infty d
z\,\frac{1}{r}\, \exp\left(\!-\frac{r}{a}\right)
=2\,\frac{Ze^2}{\beta}K_0\left(\frac{b}{a}\right),\nn
\ea
where  $\vec q$ is the momentum transfer, $(z,\vec{b})$ are the
longitudinal and transverse coordinates respectively, and $\phi(b)$
is the eikonal phase in the case of the screened Coulomb potential
with the Thomas-Fermi atom radius  $a$ and the modified Bessel
function $K_0(b/a)$.

It is convenient to introduce a two-dimensional potential $V(b)$
that appears in the Landau-Pomeranchuk-Migdal effect theory when
solving a transport equation (see Appendix A in \cite{K}):
\ba V(b)=n\int\Big[1&-&\exp(i\,\vec{q}\,\vec{ b})\Big]\left\vert
f(\vec{q})\right\vert^2d^2q\ , \\\nn
\left\vert
 f(\vec{q})\right\vert^2d^2q&=&d\sigma(q)\ .
\ea
The equation for the potential $V(b)$ can be written (after
performing the angular integration) as
\ba\label{Baier} \frac{V(b)}{2\pi n}=\int[1-J_0(q b)]d\sigma(q)\
 . \ea
Comparing this result with
\ba \frac{N_0-N(\eta)}{n\,t}=\int [1-J_0(\eta\chi)]d\sigma(\chi)\ ,
\ea
in which $N_0-N(\eta)= -\ln g(\eta)$, we obtain the similarity with
(\ref{Baier}) when accepting $qb=\eta\chi,\; q=p\eta,\; b=\chi/p,\;
p=mv$.

So the problem of deviation of the potential $V(b)$ from the Born
one $V^{\scriptscriptstyle B}(b)$
\ba \Delta V(b)&=&-\Delta_{\scriptscriptstyle
 CC}[V(b)]=
V(b)-V^{\scriptscriptstyle B}(b)=\nonumber
\\
&=&n\int d^2 x\left\{\exp\Big \{i\left[\phi\left(\vert\vec{
b}+\vec{x}\vert\right)- \phi(x)\right]\Big \}\right.\nn\\
&&\left.-\;1+\frac{1}{2}\left[\phi\left(\vert\vec{
b}+\vec{x}\vert\right)-\phi(x)\right]^2\right\}\ , \ea
where $\vec x=\gamma\vec b$, and $\gamma$ is the usual relativistic
factor of the scattered particle, is similar to our problem of
deviation of the screening angle in the eikonal approximation from
its Born value:
\ba\label{g} \Delta\big[\!-\ln
g(\eta)\big]=\Delta_{\scriptscriptstyle CC}\left[\ln
g(\eta)\right]=\frac{1}{2}\left(\chi_c\eta\right)^2\Delta_{\scriptscriptstyle
CC}\left[\ln\big(\chi_a^{\,\prime}\big)^2\right]\nn\\
=\left(\chi_c\eta\right)^2\frac{1}{2\pi}\int d^2x \left[
\left(\frac{(\vec{x}+
\vec{b})^2}{x^{\,2}}\right)^{i\xi}-1+\frac{\xi^2}{2}
\ln^2\frac{(\vec{x} +\vec{ b})^2}{x^{\,2}} \right]\nn\ea
\vspace{-5mm}
\begin{equation}=\left(\chi_c\eta\right)^2f(\xi) \end{equation}
with the Coulomb corrections $\Delta_{\scriptscriptstyle CC}\big[\ln
g(\eta)\big]\equiv \ln g(\eta)-\ln g^{\scriptscriptstyle B}(\eta)$,
$\Delta_{\scriptscriptstyle
CC}[\ln\big(\chi_a^{\,\prime}\big)]\equiv\ln\big(\chi_a^{\,\prime}\big)-\ln
\big(\chi_a^{\,\prime}\big)^{\scriptscriptstyle B}$, and
$\chi_a^{\,\prime}\equiv 1.080\,\chi_a$. The accuracy of
transformations in going from (\ref{Baier}) to (\ref{g}) coincides
with the accuracy of the eikonal approximation.

The two-dimensional integral calculated in \cite{K} turns out to be
an universal function of the Born parameter $\xi$ which is also
known as the Bethe-Maximon function:
\ba\label{summa}
f(\xi)=\xi^2\sum\nolimits_{n=1}^\infty\frac{1}{n(n^2+\xi^2)}\ . \ea

From (\ref{g}), we obtain
\ba\label{res} \Delta_{\scriptscriptstyle
CC}[\ln\big(\chi_a^{\,\prime}\big)]=f(\xi)\equiv
\Re\big[\psi(1+i\xi)\big ]+C_{\scriptscriptstyle E}\ea
with $C_{\scriptscriptstyle E}=-\psi(1)$ and the digamma function
$\psi$. Here we use the smallness of the ratios $x/a\ll 1$, $b\sim
x\ll a$ and apply the relevant asymptotes of the Bessel function
$K_0(z)=C-\ln(z/2)+O(z^2)$. The main reason of such derivation of
relations (\ref{g}) and (\ref{res}) is the significantly different
regions of contributions of the screening effects and the Coulomb
corrections. Really, the last ones play the main role in the region
of small impact parameters, where the number of atom electrons is
small and the screening effects are negligible. These results are
valid in the ultra-relativistic case considered in \cite{K}. They
can also be obtained by using the technique developed in \cite{TVG}.

In order to calculate in $\xi$ the exact absolute correction
$\Delta_{\scriptscriptstyle
CC}[\ln\big(\chi_a^{\,\prime}\big)]=f(\xi)$ and exact relative
correction $\delta_{\scriptscriptstyle CC}[\chi_a] $ to the Born
screening angle
\ba\label{del0}\delta_{\scriptscriptstyle
CC}[\chi_a^{\,\prime}]=\delta_{\scriptscriptstyle
CC}[\chi_a]=\left(\chi_a- \chi_a^{\scriptscriptstyle
B}\right)/\chi_a^{\scriptscriptstyle B}=
\exp\left[f\left(\xi\right)\right]-1, \ea
we must first calculate the values of the function
$f(\xi)=\Re\big[\psi(1+i\xi)\big ]+C_{\scriptscriptstyle E}$.  The
digamma series
\ba \psi(1+\xi)&=&1-C_{\scriptscriptstyle E}-\frac{1}{1+\xi}
+\sum\limits_{n=2}^{\infty}(-1)^{n}\big[\zeta(n-1)\big]\,\xi^{n-1},\nn
\ea
where $\zeta$ is the Riemann zeta function and $\vert \xi\vert<1$,
leads to the corresponding power series for
$\Re\big[\psi(1+i\xi)\big ]=\Re\big[\psi(i\xi)\big ]$ and $\vert
\xi\vert<2$:
\ba \Re\big[\psi(i\xi)\big ]\!=\!1\!-\!C_{\scriptscriptstyle
E}\!-\!\frac{1}{1\!+\!\xi^2}
\!+\!\sum\limits_{n=1}^{\infty}(-1)^{n\!+\!1}\big[\zeta(2n\!+\!1)\big]\,\xi^{2n}.\nn
\ea
The function
$f(\xi)=\xi^2\sum\limits_{n=1}^{\infty}[n(n^2+\xi^2)]^{-1}$ can be
represented in this cases as \cite{ryzh}
\ba\label{f(a)}
f(\xi)=1-\frac{1}{1+\xi^2}+\sum\limits_{n=1}^{\infty}(-1)^{n+1}\big[\zeta(2n+1)-1\big]\,
\xi^{2n}\nn\ea
\vspace{-5mm}
\begin{equation}
=1-\frac{1}{1+\xi^2}+0.2021\,\xi^2-0.0369\,\xi^4+0.0083\,\xi^6-\ldots
\end{equation}
The calculation results for function $f(\xi)$ (\ref{f(a)}) and the
relative Coulomb correction $\delta_{\scriptscriptstyle CC}[\chi_a]$
(\ref{del0}) at $\beta=1$ and $z=1$ are given in Table 1 (see also
Figure 1).

During our analysis, we omit systematically the contribution of an
order of $\alpha$ compared with that of an order of 1. We emphasize
that only the ultrarelativistic case is considered during our
numerical calculations, so $\beta=v/c=1$.

We can also compare (\ref{del0}) with the Moli\`{e}re
 result
 $\delta^{}_{\scriptscriptstyle M}[\chi_a]$:
\ba \label{corr3}\delta_{\scriptscriptstyle
CCM}[\delta_{\scriptscriptstyle
CC}]=\frac{\delta^{}_{\scriptscriptstyle
CC}[\chi_a]-\delta^{}_{\scriptscriptstyle M}[\chi_a] }
{\delta^{}_{\scriptscriptstyle
M}[\chi_a]}=\frac{\Delta_{\scriptscriptstyle
CCM}[\delta_{\scriptscriptstyle CC}]}{\delta^{}_{\scriptscriptstyle
M}[\chi_a]} \,. \ea

For this purpose, we rewrite (\ref{1}) as
\ba\label{del2}\delta^{}_{\scriptscriptstyle
 M}[\chi_a]&=&\left(\chi^{\scriptscriptstyle
M}_a-\chi_a^{\scriptscriptstyle B}\right)/
\chi_a^{\scriptscriptstyle B}= \sqrt{1+3.34\,\xi^2}-1\ . \ea

In order to obtain the relative difference between the  approximate
$\chi_a^{\scriptscriptstyle M}$ and exact $\chi_a$ results for the
screening angle
\begin{eqnarray}\label{angle}
\delta_{\scriptscriptstyle CCM}[\chi_a]&\equiv&\left(\chi_a-
\chi_a^{\scriptscriptstyle M}\right)/\chi_a^{\scriptscriptstyle
M}=\chi_a/\chi_a^{\scriptscriptstyle M}-1,
\end{eqnarray}
we rewrite (\ref{del0}) and (\ref{del2}) in the form
\begin{equation}
\label{folform} \delta_{\scriptscriptstyle CC}[\chi_a]+1=\chi_a/
\chi_a^{\scriptscriptstyle B}\ ,\qquad \delta^{}_{\scriptscriptstyle
M}[\chi_a]+1=\chi^{\scriptscriptstyle M}_a/
\chi_a^{\scriptscriptstyle B}\end{equation}
and obtain the expression
\begin{eqnarray}
\label{rat} \delta_{\scriptscriptstyle
CCM}[\chi_a]=\frac{\delta_{\scriptscriptstyle CC}
[\chi_a]+1}{\delta^{}_{\scriptscriptstyle M}[\chi_a]+1}-1.
\end{eqnarray}

\begin{table}
\caption{\label{}The $Z$ dependence of the Coulomb corrections and
differences defined by Eqs. (\ref{del0})--(\ref{corr3}) and
(\ref{rat})--(\ref{rel})  for $z=1$, $\beta=1$, and
$B^{\scriptscriptstyle}=8.46$.}
\begin{ruledtabular}
\begin{tabular}{rccccccccc}
$Z$~~~&$f(\xi)$~~~& $\delta_{\scriptscriptstyle
CC}[\chi_a]$&$\delta_{\scriptscriptstyle
M}[\chi_a]$&$\Delta_{\scriptscriptstyle
CCM}[\delta_{\scriptscriptstyle CC}]$&$\delta_{\scriptscriptstyle
CCM}[\delta_{\scriptscriptstyle CC}]$& $\delta_{\scriptscriptstyle
CCM}[\chi_a]$&$\delta_{\scriptscriptstyle
CC}\!\left[\overline{\theta^2}\right]$&$\Delta_{\scriptscriptstyle
CC}\left[b\right]$&$\Delta_{\scriptscriptstyle CC}\left[B\right]$\\
\hline\\[-3mm]
6~~~& 0.002~~~&0.002&0.003&$-0.001$&$-0.282$&$-0.001$&$-0.000$&$-0.002$&$-0.003$\\
13~~~&0.011~~~&0.011&0.015&$-0.004$&$-0.276$&$-0.004$&$-0.001$&$-0.011$&$-0.012$\\
22~~~&0.030~~~&0.031&0.042&$-0.011$&$-0.270$&$-0.011$&$-0.004$&$-0.030$&$-0.034$\\
28~~~&0.049~~~&0.050&0.068&$-0.018$&$-0.266$&$-0.017$&$-0.006$&$-0.049$&$-0.055$\\
42~~~&0.105~~~&0.110&0.146&$-0.036$&$-0.297$&$-0.031$&$-0.014$&$-0.105$&$-0.119$\\
50~~~&0.144~~~&0.154&0.202&$-0.047$&$-0.246$&$-0.040$&$-0.019$&$-0.144$&$-0.163$\\
73~~~&0.276~~~&0.318&0.396&$-0.078$&$-0.198$&$-0.056$&$-0.037$&$-0.276$&$-0.313$\\
78~~~&0.307~~~&0.359&0.443&$-0.084$&$-0.189$&$-0.058$&$-0.041$&$-0.307$&$-0.348$\\
79~~~&0.312~~~&0.367&0.452&$-0.085$&$-0.188$&$-0.059$&$-0.042$&$-0.355$&$-0.355$\\
82~~~&0.332~~~&0.393&0.482&$-0.089$&$-0.185$&$-0.060$&$-0.045$&$-0.332$&$-0.376$\\
92~~~&0.395~~~&0.484&0.583&$-0.099$&$-0.169$&$-0.062$&$-0.053$&$-0.395$&$-0.448$
\end{tabular}
\end{ruledtabular}
\end{table}

\begin{figure}[h!]
\begin{center}
\includegraphics[width=0.58\linewidth]{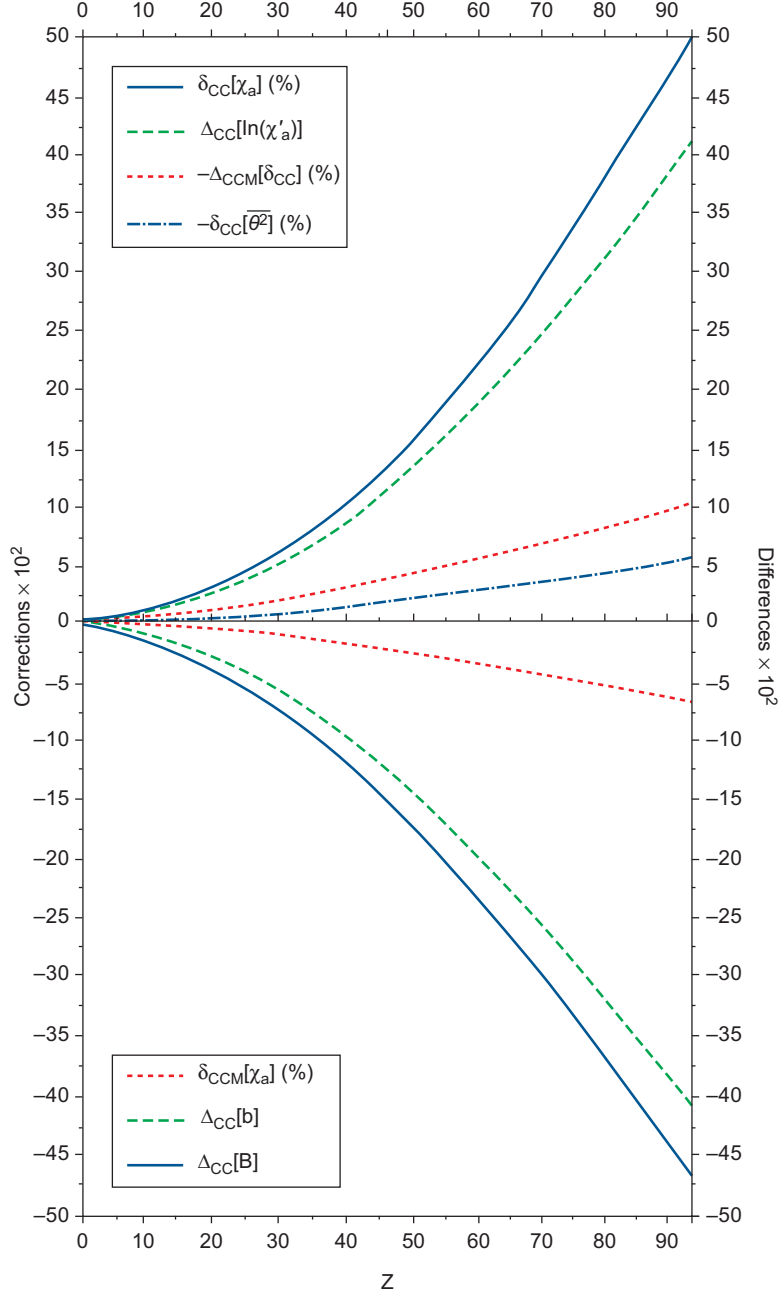}%
\caption{The $Z$ dependence of the Coulomb corrections
$\Delta_{\scriptscriptstyle CC}$, $\delta_{\scriptscriptstyle CC}$
to some parameters of the Moli\`{e}re theory and the differences
$\Delta_{\scriptscriptstyle CCM}$, $\delta_{\scriptscriptstyle CCM}$
between exact and approximate results.
 \label{fig:epsart}}
\end{center}
\end{figure}

We calculate also the Coulomb corrections to other important
parameters of the Moli\`{e}re theory. Inserting (\ref{b}) into
(\ref{B}) and differentiating the latter, we arrive at
\begin{equation}\label{limcorrec1}
\Delta_{\scriptscriptstyle CC}[b]=-f(\xi)
=\left(1-1/B^{\scriptscriptstyle
B}\right)\cdot\Delta_{\scriptscriptstyle CC}[B]\ .
\end{equation}
So $\Delta_{\scriptscriptstyle CC}[B]$ becomes
\begin{equation}\label{limco1}
\Delta_{\scriptscriptstyle CC}[B]=f(\xi)/(1/B^{\scriptscriptstyle
B}-1)\ .
\end{equation}
Accounting for $\overline{\theta^2}=\chi_c^2B$ (\ref{vartheta2}), we
get
\begin{equation}\label{CCvarthet2}
\Delta_{\scriptscriptstyle CC}\left[\overline{\theta^2}\right]\equiv
\overline{\theta^2}-\left(\overline{\theta^2}\right)^{\scriptscriptstyle
B}=\chi_c^2 \cdot\Delta_{\scriptscriptstyle CC}\left[B\right]\ .
\end{equation}
Finally, the relative Coulomb corrections can be represented as
\begin{equation}\label{rel}
\delta_{\scriptscriptstyle
CC}\left[\overline{\theta^2}\right]=\delta_{\scriptscriptstyle
CC}\left[B\right] =f(\xi)/(1-B^{\scriptscriptstyle B})\ .
\end{equation}
The $Z$ dependence of the corrections (\ref{del0}), (\ref{f(a)}),
(\ref{limcorrec1}), (\ref{limco1}), (\ref{rel}),
 and the relative differences (\ref{corr3}), (\ref{rat}) are presented in Table 1.
 Some
results from Table 1 are illustrated in Figure~\ref{fig:epsart}.

Table 1 shows that while the modulus of $\delta_{\scriptscriptstyle
CC}\left[\overline{\theta^2}\right]$ value reaches only about $5\%$
for high $Z$ targets, the maximum $\delta_{\scriptscriptstyle
CC}[\chi_a]$ value is an order of magnitude higher and amounts
approximately to 50\% for $Z=92$. It is also obvious that whereas
the relative difference $\delta_{\scriptscriptstyle CCM}$  between
exact and approximate results (\ref{corr3}) and (\ref{rat})
 varies between 17 and 28\% over the range $6\eqslantless Z
\eqslantless 92$
 for the relative Coulomb correction $\delta_{\scriptscriptstyle CC}[\chi_a]$
 to the screening angle, it reaches only about 6\% for the screening angle $\chi_a$
itself at $Z=92$. As can be seen from Table 1, modules of the
Coulomb corrections to the parameters $b$ and $B$ reach large values
for heavy target elements. So $-\Delta_{\scriptscriptstyle
CC}[B]\sim 0.45$, $-\Delta_{\scriptscriptstyle CC}[b]\sim 0.40$,
such as $\Delta_{\scriptscriptstyle
CC}[\ln\big(\chi_a^{\,\prime}\big)]\sim 0.40$ for $Z=92$. Let us
notice also that the sizes of the Coulomb corrections
$-\delta_{\scriptscriptstyle CC}\left[\overline{\theta^2}\right]$
and $-\Delta_{\scriptscriptstyle CC}[B]$, which  depend on the
parameter $B^{\scriptscriptstyle B}$, increase to $0.112$ and
$0.551$, respectively, with decreasing  $B^{\scriptscriptstyle
B}=8.46$ \cite{LPM} to minimum value $B^{\scriptscriptstyle B}=4.5$.

{\it Summary.} Within the eikonal approach, we have obtained exact
analytical results for the Coulomb corrections to the parameters
$\chi_a^{\,\prime}$, $\chi_a^{}$, $b$, $B$, and
$\overline{\theta^2}$ of the Moli\`{e}re expansion method. We
estimated numerically  these Coulomb corrections to the parameters
of the Moli\`{e}re theory for homogeneous absorbers with no energy
loss and ultra-relativistic charged projectiles
 over the range $6\eqslantless Z \eqslantless 92$ ($\beta=1$, $z=1$),
 and we found that the corrections
 $\Delta_{\scriptscriptstyle CC}[\ln\big(\chi_a^{\,\prime}\big)]$,
$\delta_{\scriptscriptstyle CC}[\chi_a]$,
 $-\Delta_{\scriptscriptstyle CC}\left[b\right]$,
 $-\Delta_{\scriptscriptstyle CC}\left[B\right]$ have large values
that increase up to $0.4$--$0.5$ for $Z\sim 95$. These large Coulomb
corrections should be taken into account in the description of high-energy
experiments with nuclear targets. The accuracy of the
Moli\`{e}re theory in determining the screening angle must also be
borne in mind.

\end{document}